\PassOptionsToPackage{table}{xcolor}
\documentclass[sigconf, nonacm]{acmart}

\newcommand\vldbdoi{XX.XX/XXX.XX}
\newcommand\vldbpages{XXX-XXX}
\newcommand\vldbvolume{14}
\newcommand\vldbissue{1}
\newcommand\vldbyear{2020}
\newcommand\vldbauthors{\authors}
\newcommand\vldbtitle{\shorttitle} 
\newcommand\vldbavailabilityurl{https://github.com/lfd/vldb26}
\newcommand\vldbpagestyle{plain} 

\newcommand{\eg}{\emph{e.g.}\xspace}
\newcommand{\ie}{\emph{i.e.}\xspace}
\newcommand{\OTHR}{Technical University of Applied Sciences Regensburg}
\newboolean{anonymous}
\setboolean{anonymous}{true}
\newboolean{diff}
\setboolean{diff}{false}

\newcommand{\etal}{\emph{et al.}\xspace}

\newcommand{\webref}[2]{\href{#1}{\textcolor{blue}{#2}}}
\newcommand{\suppweb}{\webref{https://github.com/lfd/vldb26}{reproduction package}\xspace}

\usepackage[a-2b]{pdfx}
\usepackage{mathtools}
\usepackage{amsthm}
\usepackage{amsmath}
\usepackage[group-separator={,}, group-minimum-digits={3}]{siunitx}
\usepackage{xcolor}
\usepackage{pifont}
\usepackage{circledsteps}
\usepackage{tikz}
\usetikzlibrary{calc,graphs}
\usetikzlibrary{shapes.geometric, arrows, arrows.meta}
\usetikzlibrary{matrix}
\usetikzlibrary{shapes.multipart}
\usetikzlibrary{shapes.arrows}
\usetikzlibrary{positioning}
\usetikzlibrary{backgrounds}
\usepackage[nomessages]{fp}
\usepackage{scalefnt}
\usepackage{pgfplots}
\usepackage{booktabs}
\usepackage{multirow}
\usepackage{tabularx}
\usepackage{makecell}
\usepackage{ragged2e}
\usepackage{braket}
\usepackage{xspace}
\usepackage{tabularx}
\usepackage{adjustbox}
\usepackage{ifthen}
\usepackage{enumitem} 
\usepackage[group-separator={,}, group-minimum-digits={3}]{siunitx}
\usepackage{xspace}
\usepackage[ruled,vlined,noend,linesnumbered]{algorithm2e}
\usepackage[UKenglish]{babel}
\usepackage{wrapfig}

\newtheorem{example}{Example}[section]

\tikzset{graphnode/.style={circle, text centered, minimum size=5pt, inner sep=0pt, outer sep=2pt}}
\tikzset{depthnode/.style={circle, text centered, draw=black, fill=lfd7!60, minimum size=5pt, inner sep=0pt, outer sep=2pt}}
\tikzset{procnode/.style={draw, align=center, fill=lfd1!20}, inner sep=1em}
\tikzset{reledge/.style={-Stealth}}
\tikzset{stepArrow/.style={line width=2mm, gray!50,
                           -{Triangle Cap []. Fast Triangle[] Fast Triangle[]}}}

\begin{document}
\title{Hybrid Mixed Integer Linear Programming for \\ Large-Scale Join Order Optimisation}

\author{Manuel Schönberger}
\affiliation{
\institution{\OTHR}
  \city{Regensburg}
  \country{Germany}
  }
\email{manuel.schoenberger@othr.de}

\author{Immanuel Trummer}
\affiliation{
\institution{Cornell University}
\city{Ithaca}
\state{NY}
\country{USA}
}
\email{itrummer@cornell.edu}

\author{Wolfgang Mauerer}
\affiliation{
  \institution{\OTHR}
  \institution{and Siemens AG, Technology}
  \city{Regensburg/Munich}
  \country{Germany}}
\email{wolfgang.mauerer@othr.de}

\renewcommand{\shortauthors}{Schönberger, Trummer, and Mauerer}

\begin{abstract}

Finding optimal join orders is among the most crucial steps to be performed by query optimisers. Though extensively studied in data management research, the problem remains far from solved: While query optimisers rely on exhaustive search methods to determine ideal solutions for small problems, such methods reach their limits once queries grow in size. Yet, large queries become increasingly common in real-world scenarios, and require suitable methods to generate efficient execution plans. While a variety of heuristics have been proposed for large-scale query optimisation, they suffer from degrading solution quality as queries grow in size, or feature highly sub-optimal worst-case behavior, as we will show. 

We propose a novel method based on the paradigm of \emph{mixed integer linear programming~(MILP)}: By deriving a novel MILP model capable of optimising arbitrary bushy tree structures, we address the limitations of existing MILP methods for join ordering, and can rely on highly optimised MILP solvers to derive efficient tree structures that elude competing methods. To ensure optimisation efficiency, we embed our MILP method into a \emph{hybrid framework}, which applies MILP solvers precisely where they provide the greatest advantage over competitors, while relying on more efficient methods for less complex optimisation steps. Thereby, our approach gracefully scales to extremely large query sizes joining up to 100 relations, and consistently achieves the most robust plan quality among a large variety of competing join ordering methods.

\end{abstract}

\maketitle

\pagestyle{\vldbpagestyle}
\begingroup\small\noindent\raggedright\textbf{PVLDB Reference Format:}\\
\vldbauthors. \vldbtitle. PVLDB, \vldbvolume(\vldbissue): \vldbpages, \vldbyear.\\
\href{https://doi.org/\vldbdoi}{doi:\vldbdoi}
\endgroup
\begingroup
\renewcommand\thefootnote{}\footnote{\noindent
This work is licensed under the Creative Commons BY-NC-ND 4.0 International License. Visit \url{https://creativecommons.org/licenses/by-nc-nd/4.0/} to view a copy of this license. For any use beyond those covered by this license, obtain permission by emailing \href{mailto:info@vldb.org}{info@vldb.org}. Copyright is held by the owner/author(s). Publication rights licensed to the VLDB Endowment. \\
\raggedright Proceedings of the VLDB Endowment, Vol. \vldbvolume, No. \vldbissue\ %
ISSN 2150-8097. \\
\href{https://doi.org/\vldbdoi}{doi:\vldbdoi} \\
}\addtocounter{footnote}{-1}\endgroup

\ifdefempty{\vldbavailabilityurl}{}{
\vspace{.3cm}
\begingroup\small\noindent\raggedright\textbf{PVLDB Artifact Availability:}\\
The source code, data, and/or other artifacts have been made available at \url{\vldbavailabilityurl}.
\endgroup
}

\section{Introduction}
\label{sec:introduction}

The fundamental database issue of join order optimisation remains one of the most crucial steps to be performed in query optimisation. As solution costs can vary by orders of magnitude, query optimisers must avoid suboptimal join orders that yield extremely large query execution times~\cite{Neumann.2018, Trummer.2017, schoenberger:23:pvldb}. For small queries, query optimisers rely on exhaustive search methods to obtain optimal solutions. However, as join ordering is NP-hard~\cite{Cluet1995} (even in the approximate variant~\cite{Chatterji2002a}), the performance of exhaustive search methods quickly degrades as queries grow in size. Yet, large queries remain a challenge to be addressed in real-world workloads~\cite{Dieu2009, Neumann.2018, Vogelsgesang2018a}. For such queries, query optimisers shift towards more efficient heuristic join ordering approaches, such as genetic algorithms as applied by PostgreSQL. 

However, such heuristic methods typically provide no guarantees on solution quality, and often produce extremely costly join orders. In contrast, methods such as the conventional polynomial-time IKKBZ algorithm~\cite{Ibaraki.1984, Krishnamurthy.1986} efficiently yield optimal solutions within a restricted solution space encompassing only left-deep join trees. Yet, such linear join ordering solutions can exceed optimal solution costs by orders of magnitude~\cite{Neumann.2018}, which prompts the search for scalable join ordering methods capable of identifying efficient general bushy tree solutions. To this end, Neumann and Radke have derived an adaptive join ordering approach utilising a search space linearisation technique that applies dynamic programming to obtain bushy trees based on linear IKKBZ join orders~\cite{Neumann.2018}. However, the guarantees on solution optimality provided by IKKBZ do not generally translate to the quality of bushy join trees obtained by linearisation techniques. As we will empirically show, linearised join orders frequently fail to capture ideal bushy tree structures, resulting in highly suboptimal plans. 

To address the limitations of existing join ordering methods, we propose a novel hybrid optimisation method based on the established paradigm of mixed integer linear programming~(MILP). By formulating join ordering as a MILP problem, we can rely on highly optimised MILP software solvers with decades of maturing, which renders them ideal tools for optimising large-scale problems. As a substantial benefit over most competing join ordering approaches for large-scale queries, our novel method benefits from formal guarantees on solution optimality provided by the MILP optimisation. Thereby, our method achieves remarkable robustness for extremely large queries joining up to 100 relations, as we will empirically show.

Our novel hybrid MILP method improves over the existing MILP approach for join ordering proposed by Trummer and Koch~\cite{Trummer.2017} in multiple regards. Firstly, their approach was limited to left-deep join trees, which frequently results in substantial cost overheads compared to bushy join trees. In contrast, we derive a novel MILP model capable of identifying bushy join trees. However, rather than exploring the complete bushy tree solution space, which is largely filled with highly undesirable suboptimal solutions, we maintain a lightweight MILP model size by focusing the optimisation on specific sets of bushy join trees contained within a carefully selected \emph{join tree template}. In our paper, we identify suitable templates consistently capturing highly efficient join tree structures for a large number of NP-hard tree queries.

In addition to the restriction to left-deep trees, the existing MILP method~\cite{Trummer.2017} further features only limited scalability, which prompts frequent timeouts without obtaining any solution once queries grow in size~\footnote{As assessed by Neumann and Radke~\cite{Neumann.2018}, the existing MILP method experiences timeouts without obtaining any solution once queries join 40 relations or more.}. To prevent such limitations, and to optimise scalability aptness, we embed our novel MILP method in a hybrid framework that (1) utilises MILP precisely for those parts of the join tree optimisation where the MILP solver provides the greatest benefit over competitors by identifying complex bushy tree structures, and (2) switches to more efficient join ordering alternatives for those tree portions where such methods are likely to yield optimal or near-optimal solutions. In particular, for scenarios where linear left-deep shapes constitute optimal solutions, methods like IKKBZ efficiently identify ideal plans, while relying on MILP constitutes a waste of optimisation resources in such cases. Our hybrid method achieves maximum efficiency by selecting the most suitable join ordering algorithm for each optimisation step.

\paragraph{Contributions} In summary, our research contributions are as follows:

\begin{enumerate}[nosep,left=0pt]
\item We derive a novel MILP encoding for join order optimisation, which allows the use of highly efficient MILP solvers to identify complex bushy tree structures that elude competing join ordering methods.
\item We propose a hybrid algorithm that combines our novel MILP method and complementary join ordering approaches, ensuring that resource-intensive MILP is used precisely where it provides the greatest advantage over competing approaches. We thereby substantially boost MILP model efficiency, and render our approach suitable for large-scale query optimisation.
\item We conduct an empirical analysis that compares our novel hybrid MILP approach against a wide range of competitors featuring a large variety of characteristics, including conventional dynamic programming methods, polynomial-time heuristics, greedy heuristics, as well as probabilistic algorithms.
\item We demonstrate the remarkable robustness of our hybrid MILP method for extremely large query loads, including NP-hard tree queries that join up to 100 relations. Our approach obtains optimal or near-optimal solutions for most of the 900 queries considered in our analysis, and avoids the worst-case behavior of other join ordering methods, which frequently exceed best solution costs by orders of magnitude. In contrast to competitors, our hybrid MILP algorithm thus scales gracefully alongside increasing query sizes, and maintains high solution quality even for extremely large problem loads.
\end{enumerate}

The remainder of this paper is structured as follows: We begin by outlining our considered join ordering model in Sec.~\ref{sec:join_ordering_model}. We present our novel MILP encoding in Sec.~\ref{sec:milp_encoding}, and detail our hybrid framework combining MILP and complementary methods in Sec.~\ref{sec:hybrid_algorithm}. We present our experimental results in Sec.~\ref{sec:experimental_analysis}, and discuss related work in Sec.~\ref{sec:related_work}. Finally, we conclude in Sec.~\ref{sec:conclusion}.

\section{Join Ordering Model}
\label{sec:join_ordering_model}
In this section, we discuss the fundamentals of join order optimisation and our join ordering model. We begin by outlining the problem input in Sec.~\ref{sec:query_graph}, and consider the general join ordering solution space, as well as commonly applied search space restrictions, in Sec.~\ref{sec:join_tree}. Finally, we discuss our considered cost function in Sec.~\ref{sec:cost_function}.

\subsection{Query Graph} 
\label{sec:query_graph}

The input to the join ordering problem is given by a \emph{query graph} $Q=(V,E)$, where nodes represent base relations, and edges represent join predicates~\cite{Moerkotte.2020}. Each node $v_k \in V$ is labeled by the cardinality $n_k$ for relation $k$, while each join predicate is labeled by the join selectivity $0 < f_{kl} \leq 1$ for joining relations $k$ and $l$. 

In contrast to some competing join ordering methods, which require certain query graph properties such as connectivity or an absence of cycles, as presupposed, \eg, by the conventional IKKBZ algorithm~\cite{Ibaraki.1984, Krishnamurthy.1986}, our MILP model as proposed in Sec.~\ref{sec:milp_encoding} is not restricted to specific graph shapes or properties. 

\subsection{Join Tree} 
\label{sec:join_tree}

Each solution to the join ordering problem is given by a \emph{join tree}, where tree nodes correspond to join operations, with the exception of leaf nodes, which represent joined base relations. The general size of the join ordering solution space is hence given by all possible tree shapes that can be expressed by individual solutions. To handle the extremely large solution space, and to enhance the exploration efficiency, most join ordering methods only consider solutions of certain properties. 

\paragraph{Cross Products} Most prominently, join ordering methods often exclude \emph{cross product operations}, \ie, joining base relations that do not feature a shared join predicate. While these operations typically produce very costly results, and can thus often be safely disregarded, cross products can be optimal in rare cases~\cite{Gubichev2015}. Our novel MILP model explores an extended solution space including cross-product operations.

\paragraph{Tree Shape} As a further, commonly applied solution space restriction, many join ordering methods, including the conventional IKKBZ method~\cite{Ibaraki.1984, Krishnamurthy.1986}, only consider \emph{linear}, or \emph{left-deep} join trees. Depending on the query graph shape, this search space restriction can be very effective: For star queries, where the central relation constitutes a crucial operand to be featured by all joins, optimal solutions accordingly correspond to left-deep join trees. Yet, for other, more general query graph shapes, enforcing left-deep tree shapes can severely degrade solution quality by orders of magnitude~\cite{Neumann.2018}. To address this limitation of the existing MILP encoding for join ordering by Trummer and Koch~\cite{Trummer.2017}, our novel encoding allows the specification of arbitrary tree structures to be optimised.

\subsection{Cost Function}
\label{sec:cost_function}

Finally, each join tree is evaluated based on a cost function. In our paper, we consider the conventional cost function $\mathit{c_{out}}$, which sums over the sizes of all intermediate join results~\cite{Moerkotte.2020}. For a list of join-operand assignments represented by a join tree $\mathit{T}$, costs are derived as
\begin{equation}
    \label{eqn:c_out}
    \mathit{c_{out}(T)} = \sum_{\mathit{Op_j} \in \mathit{T}} \left( \prod_{r \in \mathit{Op_j}}{n_r} \cdot \prod_{(r_k, r_l) \subseteq \mathit{Op_j}} \mathit{f_{kl}} \right),
\end{equation}
where $\mathit{Op_j}$ denotes a list of relations $r$ used as operands for join $j$. 

Note that we do not consider the costs of the final (root) join in our cost calculation, as this cost value is invariant w.r.t. individual join ordering solutions and join trees, and would be merely added as a constant offset to the costs of any solution. Thereby, we avoid cases where the cost offset yielded by the final join equalises the solutions yielded by varying algorithms that otherwise vastly differ in costs. This may occur if final join costs substantially exceed the accumulated costs contributed by the remaining intermediate joins.


\section{MILP Encoding}
\label{sec:milp_encoding}
\begin{figure}[htbp]
  \centering
\tikzstyle{graphnode} = [circle, text centered, draw=black, minimum size=10pt, fill=blue!60, inner sep=0pt, outer sep=2pt]
\tikzstyle{depthnode} = [circle, text centered, draw=black, fill=blue!60, minimum size=5pt, inner sep=0pt, outer sep=2pt]
\begin{tikzpicture}[node distance=1cm]
    	   \node (node0) [graphnode] {\textcolor{white}{$i$}};
        \node (node1) [graphnode]  at ([graphnode, shift=({240:1 cm})]node0) {\textcolor{white}{$j$}};
        \node (node2) [graphnode]  at ([graphnode, shift=({240:1 cm})]node1) {\textcolor{white}{$l$}};
        \node (node5) [graphnode]  at ([graphnode, shift=({300:1 cm})]node0) {\textcolor{white}{$k$}};

        \node (node6) [graphnode, draw=white, fill=white]  at ([graphnode, shift=({270:2.5cm})]node0) {(a)};

        \draw [reledge] (node1) -- (node0);
        \draw [reledge] (node2) -- (node1);
        \draw [reledge] (node5) -- (node0);
        
        \end{tikzpicture}%
\hspace{2em}
\tikzstyle{graphnode} = [circle, text centered, draw=black, minimum size=10pt, fill=blue!60, inner sep=0pt, outer sep=2pt]
\tikzstyle{relationnode} = [circle, text centered, draw=black, minimum size=10pt, fill=teal!60, inner sep=0pt, outer sep=2pt]
\begin{tikzpicture}[node distance=1cm]
    	   \node (node0) [graphnode] {\textcolor{white}{$i$}};
        \node (node1) [graphnode]  at ([graphnode, shift=({240:1 cm})]node0) {\textcolor{white}{$j$}};
        \node (node2) [graphnode]  at ([graphnode, shift=({240:1 cm})]node1) {\textcolor{white}{$l$}};
        \node (node3) [relationnode]  at ([graphnode, shift=({300:1 cm})]node1) {\textcolor{white}{C}};
        \node (node4) [relationnode]  at ([graphnode, shift=({240:1 cm})]node2) {\textcolor{white}{A}};
        \node (node6) [relationnode]  at ([graphnode, shift=({300:1 cm})]node2) {\textcolor{white}{B}};
        \node (node5) [relationnode]  at ([graphnode, shift=({300:1 cm})]node0) {\textcolor{white}{D}};

        \node (node7) [graphnode, draw=white, fill=white]  at ([graphnode, shift=({270:3.2cm})]node0) {(b)};

        \draw [reledge] (node1) -- (node0);
        \draw [reledge] (node2) -- (node1);
        \draw [reledge] (node5) -- (node0);
        \draw [reledge] (node3) -- (node1);
        \draw [reledge] (node4) -- (node2);
        \draw [reledge] (node6) -- (node2);
        
        \end{tikzpicture}%
\hspace{2em}
\tikzstyle{graphnode} = [circle, text centered, draw=black, minimum size=10pt, fill=blue!60, inner sep=0pt, outer sep=2pt]
\tikzstyle{relationnode} = [circle, text centered, draw=black, minimum size=10pt, fill=teal!60, inner sep=0pt, outer sep=2pt]
\begin{tikzpicture}[node distance=1cm]
    	   \node (node0) [graphnode] {\textcolor{white}{$i$}};
        \node (node1) [graphnode]  at ([graphnode, shift=({235:1 cm})]node0) {\textcolor{white}{$j$}};
        \node (node2) [relationnode]  at ([graphnode, shift=({250:1 cm})]node1) {\textcolor{white}{A}};
        \node (node3) [relationnode]  at ([graphnode, shift=({290:1 cm})]node1) {\textcolor{white}{B}};
        \node (node5) [graphnode]  at ([graphnode, shift=({305:1 cm})]node0) {\textcolor{white}{$k$}};
        \node (node4) [relationnode]  at ([graphnode, shift=({250:1 cm})]node5) {\textcolor{white}{C}};
        \node (node6) [relationnode]  at ([graphnode, shift=({290:1 cm})]node5) {\textcolor{white}{D}};

        \node (node7) [graphnode, draw=white, fill=white]  at ([graphnode, shift=({270:2.5cm})]node0) {(c)};

        \draw [reledge] (node1) -- (node0);
        \draw [reledge] (node2) -- (node1);
        \draw [reledge] (node5) -- (node0);
        \draw [reledge] (node3) -- (node1);
        \draw [reledge] (node4) -- (node5);
        \draw [reledge] (node6) -- (node5);

        \end{tikzpicture}%
  \captionof{figure}{Illustrating our approach based on a tree template (a) featuring joins $i$, $j$, $k$, $l$, to join the four relations A, B, C and D. The template encompasses both a left-deep join tree (b), and a balanced join tree (c).}\label{fig:example_template}
\end{figure}
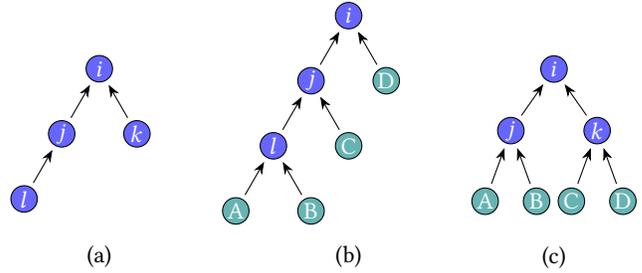

In this section, we present our novel MILP encoding that allows for the optimisation of bushy join tree structures, thus improving over the existing MILP model by Trummer and Koch~\cite{Trummer.2017}, which is limited to left-deep join trees. Rather than exploring the complete bushy join tree solution space, our model operates based on arbitrary \emph{join tree templates}: We allow the MILP optimiser to choose between varying join trees \emph{encompassed} by a specified template. Fig.~\ref{fig:example_template}~(a) illustrates a template featuring four joins $i$, $j$, $k$, and $l$ for joining four relations A, B, C and D. Given this template join arrangement, the MILP optimiser can either select a left-deep join tree as shown in Fig.~\ref{fig:example_template}~(b), or the balanced tree variant featured in Fig.~\ref{fig:example_template}~(c).

Restricting the MILP optimisation to select join trees encoded via a tree template allows us to maintain a high modeling efficiency and algorithmic performance, which constitute essential properties for our goal of large-scale join order optimisation. In contrast, MILP models for the unrestricted bushy join tree solution space would require both, a substantial encoding overhead in both variables and constraints, as well as increased search complexity that degrades performance. Naturally, the optimisation quality of our approach rests on the selection of suitable templates that capture optimal or near-optimal tree structures. We will discuss our template selection approach in Sec.~\ref{sec:hybrid_algorithm}, and empirically demonstrate its aptness in our experimental analysis in Sec.~\ref{sec:experimental_analysis}.

In the following, we discuss each MILP encoding step in detail. We explain the encoding process for join operators in Sec.~\ref{sec:join_operators}, operand relations in Sec.~\ref{sec:operand_relations}, and finally, the cost calculation in Sec.~\ref{sec:cost_calculation}. We illustrate each encoding step based on a running example for joining four relations A, B, C and D, using the join tree template shown in Fig.~\ref{fig:example_template}. Table~\ref{tab:variable_overview} and Table~\ref{tab:constraint_overview} respectively provide an overview on all variables and constraints used in our MILP model.

\newcommand{\tabitem}{~~\llap{\textbullet}~~}
\begin{table}[htbp]
	\centering
	\caption{Overview of all variables, their semantics and required amounts for queries joining $R$ relations and $P$ predicates, using a join template with $J$ joins, and $T$ threshold values for the cost calculation.}
        \rowcolors{2}{gray!10}{white}
	\begin{tabular}{llrr}  
		\toprule
		Var\({^\text{\underline{s}}}\) & Semantics & \# Variables \\
		\midrule
	 $\mathit{ja_{j}}$ & Is \emph{join} j  \emph{active}? & 
		    $J$ \\
    $\mathit{roj_{rj}}$ & Is \emph{relation} r an \emph{operand} for \emph{join} j? & $RJ$ \\
    $\mathit{paj_{pj}}$ & Is \emph{predicate} p \emph{applicable} for \emph{join} j? & $PJ$ \\
    $\mathit{trj_{tj}}$ & Is \emph{threshold} $\theta_t$ \emph{reached} by \emph{join} j? & $TJ$ \\
		\bottomrule\rowcolor{white}
	\label{tab:variable_overview}
	\end{tabular}\vspace*{-2em}
\end{table}

\begin{table*}[htbp]
	\centering
	\caption{Overview of all constraints, their semantics, and encodings used by our novel MILP model.}
        \rowcolors{2}{gray!10}{white}
	\begin{tabular}{lll}  
		\toprule
		 & Semantics & Encoding \\
		\midrule
	 (A) & Enforce correct number of active joins for a query joining $R$ relations. & 
		    $\sum_{j=1}^J \mathit{ja_j} = R-1$ \\
    (B) & Enforce join tree connectivity between join $i$ and its successor $j$. & $\mathit{ja_i} \leq \mathit{ja_j}$ \\
    (C) & Enforce correct number of operands for each join $j$. & $\sum_r^R \mathit{roj_{rj}} = 2 \cdot \mathit{ja_j} + \sum_{i \in \mathit{Pred(j)}} \mathit{ja_i}$ \\
    (D) & Ensure continuity of joined operands for a join $j$ and its successor $i$. & $\mathit{roj_{rj}} \leq \mathit{roj_{ri}}$ \\
    (E) & Prevent assignment of operands to inactive joins. & $\mathit{roj_{rj}} \leq \mathit{ja_{j}}$ \\
    (F) & Prevent conflicting operand assignments for joins $i$ and $j$ with  a shared successor. & $\mathit{roj_{ri}} + \mathit{roj_{rj}} \leq 1$ \\
    (G) & Prevent invalid join predicates for each join $j$ and predicate $p$. & $\mathit{pao_{pj}} \leq \mathit{roj_{Rel_1(p)j}}, \mathit{pao_{pj}} \leq \mathit{roj_{Rel_2(p)j}}$ \\
    (H) & Enforce cost threshold value activation for each join $j$ and threshold $\theta_t$. & $\mathit{LogIntCard(j)} - \mathit{trj_{tj}} \cdot \infty \leq \log(\theta_t)$ \\
		\bottomrule\rowcolor{white}
	\label{tab:constraint_overview}
	\end{tabular}\vspace*{-2em}
\end{table*}

\subsection{Encoding Joins Operators}
\label{sec:join_operators}

We begin by discussing variables and constraints pertaining to the \emph{join operators} included in the join tree template. The template contains an arbitrary number of joins, to allow the selection of varying join trees formed by specific subsets of join operators, while \emph{excess joins} not included in the selected tree will remain unused. Accordingly, we require variables indicating which among the included joins are \emph{actively used} and form the join tree: Let the binary variable $\mathit{ja_{j}}$ (\emph{Join is Active}), introduced for each join $j$, indicate whether $j$ is active.  Further, we require constraints that enforce the creation of \emph{valid join trees}, by (A) ensuring the correct \emph{number} of \emph{active joins}, and by (B) ensuring each \emph{intermediate join} eventually \emph{connects} to the \emph{root} join. 

\paragraph{Constraint (A)} For condition (A), we add the constraint (A) $\sum_{j=1}^J \mathit{ja_j} = R-1$, where $J$ denotes the total number of joins in the template, and $R$ denotes the number of base relations to be joined. Thus, we enforce that the number of active joins correctly corresponds to $R-1$, such that all base relations can be joined, and any excess join beyond this bound must remain inactive.

\begin{example}
\label{example:constr_a}
\looseness-1 To illustrate each step of our MILP encoding, consider an example problem joining four relations $A, B$, $C$, and $D$, using the join tree template depicted in Fig.~\ref{fig:example_template}, with joins $i$, $j$, $k$, $l$. For all joins, we add the corresponding join variables $\mathit{ja_i}$, $\mathit{ja_j}$, $\mathit{ja_k}$ and $\mathit{ja_l}$. Our model must ensure that a MILP optimiser only selects those variable configurations that form valid join trees.

We begin our example by considering the effect of adding constraint (A) discussed above, which enforces the correct number of active joins. Since $R=4$, $\mathit{ja_i}+\mathit{ja_j}+\mathit{ja_k}+\mathit{ja_l}=R-1=3$ must hold. Thus, the optimiser may select $\mathit{ja_i}=\mathit{ja_j}=\mathit{ja_l}=1$ and $\mathit{ja_k}=0$, thereby forming a left-deep tree, or $\mathit{ja_i}=\mathit{ja_j}=\mathit{ja_k}=1$ and $\mathit{ja_l}=0$, which forms a balanced tree. These two variants exhaust all valid tree formations possible for join ordering problems featuring four base relations. However, as our set of join operator constraints is still incomplete, the optimiser may, for instance, instead select the variable configuration $\mathit{ja_j}=\mathit{ja_k}=\mathit{ja_l}=1$ and $\mathit{ja_i}=0$, which corresponds to an incomplete tree with a missing root join.
\end{example}

\paragraph{Constraint (B)} To prevent invalid variable configurations as featured in Example~\ref{example:constr_a}, we next consider constraints that enforce join tree connectivity. To this end, we rely on MILP \emph{implication encodings}, \ie, a constraint type that enforces the relationship $a \implies b$ for two binary variables $a$ and $b$. In MILP, this relationship can be enforced by adding the constraint $a \leq b$. As such, for each join $j$ in our template, and for each join $i$ directly preceding $j$ in accordance with our template, we add the constraint (B) $\mathit{ja_i} \leq \mathit{ja_j}$, which enforces join $j$ to be activated if any direct predecessor join $i$ is active: If $\mathit{ja_i}=1$ and $\mathit{ja_j}=0$, we obtain $1 \leq 0$, which violates the constraint. We will rely on similar implication encodings for other variable types discussed below.

\begin{example}
\label{example:constr_b}
\looseness-1 (cont'd) We continue our example problem joining four relations $A, B$, $C$ and $D$, using the join tree template depicted in Fig.~\ref{fig:example_template}, with joins $i$, $j$, $k$, $l$. 

We now consider the effect of constraint (B) on the variable configuration $\mathit{ja_j}=\mathit{ja_k}=\mathit{ja_l}=1$ and $\mathit{ja_i}=0$, which forms an incomplete tree without a root join as discussed in Example~\ref{example:constr_a}. Since join $i$ is preceded by both $j$ and $k$, we add constraints $\mathit{ja_j} \leq \mathit{ja_i}$ and $\mathit{ja_k} \leq \mathit{ja_i}$. However, given the variable configuration above, the optimiser encounters constraint violations, as $\mathit{ja_j} = 1 \leq 0 = \mathit{ja_i}$, and $\mathit{ja_k} = 1 \leq 0 = \mathit{ja_i}$. Accordingly, constraint (B) prevents the optimiser from choosing this variable configuration, or any other configuration representing a disconnected tree. In contrast, for the valid left-deep tree variant, no violations occur, as we obtain $\mathit{ja_l}=1 \leq 1=\mathit{ja_j}$, and $\mathit{ja_j}=1 \leq 1=\mathit{ja_i}$. The same holds, mutatis mutandis, for the balanced tree variant. 
\end{example}

By adding both constraints (A) and (B), we have ensured that the selected join operators form valid join trees encompassed by the tree template.

\subsection{Encoding Operand Relations}
\label{sec:operand_relations}

Having completed our MILP encoding for join operators, we next consider the \emph{join operands}, \ie, the base relations processed by each join: Let the binary variable $\mathit{roj_{rj}}$ (\emph{Relation is Operand for Join}), introduced for each relation $r$ and join $j$, indicate whether $r$ is an operand for $j$. As before, we require a set of constraints to ensure valid assignments of join operands. In addition to constraints (A) and (B) for join operators, we must  
\begin{itemize}[nosep,left=0pt]
    \item ensure correct operands amounts for each join (C),
    \item ensure continuity of operands (D),
    \item prevent the assignment of operands to inactive joins (E), and
    \item prevent conflicting operand assignments (F).
\end{itemize}
In the following, we discuss each of these conditions in detail.

\paragraph{Constraint (C)} We begin with condition (C), which requires the \emph{correct number of operands} $n_j$ for each join operator $j$. In the straightforward scenario of joining two base relations, $n_j=2$. However, since joins may moreover process intermediate results produced by predecessor joins, we must further consider the \emph{number of direct or indirect predecessor joins} $|\mathit{Pred(j)}|$ for $j$. In particular, we require $n_j=2+|\mathit{Pred(j)}|$ operands for any join $j$. However, rather than accounting for each join included in the join tree template, we must only consider those joins that have been activated by the optimiser, including join $j$ or any of its potential predecessors. Accordingly, for each join $j$, we add the constraint (C) $\sum_r^R \mathit{roj_{rj}} = 2 \cdot \mathit{ja_j} + \sum_{i \in \mathit{Pred(j)}} \mathit{ja_i}$.

\begin{example}
\label{example:constr_c}
\looseness-1 (cont'd) We continue our example problem joining four relations $A, B$, $C$ and $D$, using the join tree template depicted in Fig.~\ref{fig:example_template}, with joins $i$, $j$, $k$, $l$. We assume the optimiser has selected the balanced tree variant, expressed by the variables $\mathit{ja_i}=\mathit{ja_j}=\mathit{ja_k}=1$ and $\mathit{ja_l}=0$. 

Firstly, for each join and relation pair, we add the corresponding $\mathit{roj}$ variable, to indicate the assignment of relations to join operators. Next, we add the constraint type $(C)$ for each join. For the active join $j$, the optimiser then selects two operands, since $\sum_r^R \mathit{roj_{rj}} = 2 \cdot \mathit{ja_j} + \mathit{ja_l} = 2 + 0 = 2$, given the inactive join $l$. The same applies to join $k$, which features no predecessor joins in the template. Finally, further processing the intermediate results produced by joins $j$ and $k$, the optimiser selects four operands for the root join $i$, as $\sum_r^R \mathit{roj_{ri}} = 2 \cdot \mathit{ja_i} + \mathit{ja_j} + \mathit{ja_k} = 2 + 1 + 1 = 4$.
\end{example}

While our MILP encoding so far ensures the correct number of operands for each join, the specific assignment of operands otherwise remains arbitrary, which allows for invalid solutions as discussed in Example~\ref{example:constr_d} below. We thus continue by considering constraints to ensure a valid assignment of particular operands. 

\paragraph{Constraints (D) and (E)} Following the definition of join order optimisation, our MILP model must ensure continuity of joined operands, \ie, once a relation $r$ has been selected as an operand for join $j$, $r$ must moreover be featured by all joins succeeding $j$. Expressed in variables, we must enforce $\mathit{roj_{rj}} \implies \mathit{roj_{ri}}$ for all join pairs $(j, i)$, where $i$ directly succeeds $j$. Relying on the same MILP implication encoding used for constraint (B) above, we accordingly add constraint (D) $\mathit{roj_{rj}} \leq \mathit{roj_{ri}}$ for all such join pairs featured in the template. In a similar manner, we can prevent the assignment of relations to inactive joins by adding a constraint (E) $\mathit{roj_{rj}} \leq \mathit{ja_{j}}$, which enforces $\mathit{roj_{rj}} \implies \mathit{ja_{j}}$.

\begin{example}
\label{example:constr_d}
\looseness-1 (cont'd) We continue our example problem joining four relations $A, B$, $C$ and $D$, using the join tree template depicted in Fig.~\ref{fig:example_template}, with joins $i$, $j$, $k$, $l$. To illustrate operand continuity, we assume the optimiser has selected the left-deep tree variant, expressed by the variables $\mathit{ja_i}=\mathit{ja_j}=\mathit{ja_l}=1$ and $\mathit{ja_k}=0$. 

For the left-deep tree variant, we require two, three and four operands for joins $l$, $j$ and $i$ respectively, in accordance to constraint (C). The optimiser may select relations $A$ and $B$ for join $l$, setting $\mathit{roj_{Al}}=\mathit{roj_{Bl}}=1$. However, without the effect of constraint (D), the optimiser may further choose the variable configuration $\mathit{roj_{Bj}}=\mathit{roj_{Cj}}=\mathit{roj_{Dj}}=1$ and $\mathit{roj_{Aj}}=0$. Hence, relations $B$, $C$ and $D$ are selected for join $j$, disregarding the requirement to process relation $A$ as part of the intermediate result produced by $l$. However, by adding constraint (D), the configuration results in the violation $\mathit{roj_{Al}}=1 \leq 0 = \mathit{roj_{Aj}}$, and is accordingly rendered invalid. Thus, relation $A$ must be selected as an operand for join $j$, if it is initially joined by $l$.
\end{example}

\paragraph{Constraint (F)} For purely left-deep join trees, the constraints added thus far are sufficient to ensure solution validity. However, for any bushy tree structure supported by the template, we require one additional constraint preventing conflicting operand assignments. Specifically, for any join featuring two direct predecessor joins $(i, j)$, there must be no overlaps between the sets of operands assigned to $i$ and $j$ respectively. Hence, for all such join pairs $(i, j)$, and for each relation $r$, we add the constraint (F) $\mathit{roj_{ri}} + \mathit{roj_{rj}} \leq 1$, which is violated when $r$ is assigned to both joins $i$ and $j$.

\begin{example}
\label{example:constr_f}
\looseness-1 (cont'd) We continue our example problem joining four relations $A, B$, $C$ and $D$, using the join tree template depicted in Fig.~\ref{fig:example_template}, with joins $i$, $j$, $k$, $l$. To illustrate the prevention of conflicting operand assignments, we assume the optimiser has selected the balanced tree variant, expressed by the variables $\mathit{ja_i}=\mathit{ja_j}=\mathit{ja_k}=1$ and $\mathit{ja_l}=0$. 

Following Example~\ref{example:constr_c}, we respectively require two operands for either join $j$ and $k$ with mutual direct successor $i$. Without the effect of constraint (F), the optimiser may activate the variables $\mathit{roj_{Aj}}=\mathit{roj_{Bj}}=1$ for join $j$, and $\mathit{roj_{Ak}}=\mathit{roj_{Ck}}=1$ for join $k$. However, this implies relation $A$ is joined by both joins $j$ and $k$, and thus corresponds to an invalid join ordering solution. By adding constraint (F), the solver encounters the violation $\mathit{roj_{Aj}}+\mathit{roj_{Ak}} = 1 + 1 = 2 \leq 1$, and hence correctly identifies such conflicting operand assignments as invalid.
\end{example}

With constraints (A)-(F) in place, we have ensured both, valid join tree formations and valid operand assignments, and thus obtain valid join ordering solutions when processing our MILP model.

\subsection{Cost Calculation}
\label{sec:cost_calculation}

As a final step, we must encode the costs for each join ordering solution, to allow the optimiser to determine those solutions that minimise solution costs. As described in Sec.~\ref{sec:join_ordering_model}, we consider the conventional cost function $\mathit{c_{out}}$ featured in Eqn.~\ref{eqn:c_out}, which evaluates solutions based on accumulated intermediate result sizes, relying on base relation cardinalities and join predicate selectivities. While our encoding thus far accounts for the former, we have yet to encode join predicate applicability. Following the existing MILP encoding for left-deep trees~\cite{Trummer.2017}, we introduce a variable $\mathit{pao_{pj}}$, indicating whether predicate $p$ is applicable for join $j$. 

\paragraph{Constraint (G)} We may only allow the use of a join predicate $p$ for a join $j$ if both of its associated relations $\mathit{Rel_1(p)}$ and $\mathit{Rel_2(p)}$ are present as operands for $j$. Thus, we add the constraints (G) $\mathit{pao_{pj}} \leq \mathit{roj_{Rel_1(p)j}}, \mathit{pao_{pj}} \leq \mathit{roj_{Rel_2(p)j}}$ for each predicate $p$ and join $j$.

\begin{example}
\label{example:constr_g}
(cont'd) We continue our example problem joining four relations $A, B$, $C$ and $D$. We now assume join predicates $p_1$ for relations $(A, B)$, $p_2$ for $(B, C)$, and $p_3$ for $(C, D)$. Accordingly, for each predicate and join, we add the corresponding $\mathit{pao}$ variable to indicate predicate use. 

We assume the optimiser has selected the left-deep join tree as depicted in Fig.~\ref{fig:example_template}~(b). To reduce costs via predicate selectivities (as discussed below), the MILP optimiser seeks to activate as many $\mathit{pao}$ variables as possible. For instance, the optimiser may correctly apply predicate $p_1$ for join $l$ by activating $\mathit{pao_{1l}}$: As both associated relations $A$ and $B$ are used as operands for join $l$, none of the constraints $\mathit{pao_{1l}} \leq \mathit{roj_{Al}}$ and $\mathit{pao_{1l}} \leq \mathit{roj_{Bl}}$ are violated. In contrast, further activating $\mathit{pao_{2l}}$ results in a violation, as the associated relation $C$ is not an operand for join $l$. The optimiser may only apply $p_2$ after the succeeding join $j$, which adds the missing relation $C$.
\end{example}

Having encoded both operand types used by $\mathit{c_{out}}$, we next consider the involved operations. Crucially, $\mathit{c_{out}}$ relies on product operations, which are not supported by the MILP formalism. To circumvent this issue, we can reapply the cost calculation scheme proposed by Trummer and Koch~\cite{Trummer.2017} for the existing left-deep MILP model: By replacing all input cardinalities and predicate selectivities with their logarithmic values, we can exploit the product rule for logarithms, and thereby substitute sums for the required product operations. Finally, as illustrated below, we may then approximate the original non-logarithmic costs by introducing an arbitrary set of \emph{threshold values} $T$. We summarise their approach in the following, and refer to Ref.~\cite{Trummer.2017} for further details. 

Let the binary variable $\mathit{trj_{tj}}$ (\emph{Threshold is Reached by Join}), introduced for each join $j$ and threshold value $\theta_t \in T$, indicate whether the logarithmic intermediate result size $\mathit{LogIntCard(j)}$ produced by $j$ exceeds $\log{\theta_t}$. Relying on the product rule for logarithms, we obtain $\mathit{LogIntCard(j)}=\sum_{r=1}^R \mathit{LogCard(r)}\mathit{roj_{rj}}+ \sum_{p=1}^P \mathit{LogSel(p)}\mathit{paj_{pj}}$, where $\mathit{LogCard(r)}$ and $\mathit{LogSel(p)}$ give the logarithmic cardinality for a relation $r$ and the logarithmic selectivity for a join predicate $p$ respectively.

\paragraph{Constraint (H)} To approximate non-logarithmic costs, we further introduce the constraint (G) $\mathit{LogIntCard(j)} - \mathit{trj_{tj}} \cdot \infty \leq \log(\theta_t)$ for every threshold value $\theta_t$ and join $j$, where $\infty$ is a sufficiently large constant to ensure validity if $\mathit{trj_{tj}}=1$. Hence, if $\mathit{LogIntCard(j)} > \log(\theta_t)$, $\mathit{trj_{tj}}$ must be activated in order to obtain a valid MILP solution. Finally, we specify the approximated $\mathit{c_{out}}$ cost value as our MILP objective function: $\sum_{t=1}^T\sum_{j=1}^{J-1} \mathit{trj_{tj}}\theta_t$.

\begin{example}
\label{example:constr_h}
(cont'd) We complete our example problem joining four relations $A, B$, $C$ and $D$ by demonstrating the cost approximation. We assume the optimiser has selected the left-deep tree depicted in Fig.~\ref{fig:example_template}~(b). We further assume intermediate cardinalities $\mathit{IntCard(l)}=10$, $\mathit{IntCard(j)}=100$ and $\mathit{IntCard(i)}=\num{1000}$ for joins $l$, $j$ and $i$ respectively, with overall $\mathit{c_{out}}$ costs given by $\mathit{IntCard(l)}+\mathit{IntCard(j)}+\mathit{IntCard(i)}=10+100+\num{1000}=\num{1110}$.

To enable $\mathit{c_{out}}$ cost calculation in MILP, the solver can rely on sum operations to derive the corresponding logarithmic intermediate cardinalities $\mathit{LogIntCard(l)}=1$, $\mathit{LogIntCard(j)}=2$ and $\mathit{LogIntCard(i)}=3$. Next, to approximate the actual non-logarithmic $\mathit{c_{out}}$ costs, we consider threshold values $\theta_1=10$ and $\theta_2=100$, and introduce the corresponding $\mathit{trj}$ variables for each threshold value and join. To minimise costs, the MILP optimiser seeks to leave $\mathit{trj}$ variables inactive. While this yields no constraint violation for join $l$, we obtain $\mathit{LogIntCard(j)} - \mathit{trj_{1j}} \cdot \infty = 2 - \mathit{trj_{1j}} \cdot \infty \leq 1 = \log(\theta_1)$ for constraint (H). Thus, to satisfy the constraint, $\mathit{trj_{1j}}$ must be activated, adding the non-logarithmic cost value $\theta_1=10$ in accordance in our objective function. Likewise, both logarithmic thresholds are exceeded by $\mathit{LogIntCard(i)}$, further adding non-logarithmic costs $\theta_1+\theta_2=110$. Accordingly, we obtain approximated $\mathit{c_{out}}$ costs of $120$~\footnote{For simplicity, we have not considered the issue of overlapping threshold values in our example: As exceeding $\theta_2$ also implies exceeding $\theta_1$, the latter is added twice for join $i$. To avoid this issue, we may reduce $\theta_2$ by $\theta_1$. However, not accounting for such overlaps does not influence the actual MILP optimisation, since the resulting cost overheads correspond to mere cost offsets that do not impact solution ranking.}. Clearly, threshold choice in our example fails to accurately capture actual $\mathit{c_{out}}$ costs $\num{1110}$, illustrating the importance of a suitable threshold value selection.
\end{example}

In Example~\ref{example:constr_h}, we have shown the importance of choosing suitable threshold values for the optimisation process. While optimisation accuracy increases alongside a more plentiful choice of thresholds, algorithmic efficiency deteriorates with increasing MILP model size. Thresholds should hence be carefully selected, to strike an optimal balance between approximation quality and optimisation efficiency. In the following section, we discuss how our hybrid method, which uses MILP in conjunction with complementary join ordering approaches, provides valuable information on threshold selection.  

\section{Hybrid MILP Method}
\label{sec:hybrid_algorithm}
So far, we have discussed our novel MILP encoding that overcomes the left-deep search space limitation of the original MILP method by Trummer and Koch~\cite{Trummer.2017}. Yet, in addition to expanding the solution scope to non-linear trees, we must address further issues pertaining to the MILP paradigm itself, and thereby afflicting both, the original left-deep model as well as our novel MILP method. While MILP constitutes a powerful tool to explore complex solution spaces while providing guarantees on solution quality, its scalability aptness highly depends on the model efficiency. For the original left-deep MILP method, optimisers have been reported to fail in obtaining solutions within 60s once queries join 40 relations or more~\cite{Neumann.2018}. However, finding robust join ordering methods capable of reliably identifying efficient, complex tree structures even for large queries is precisely our main motivation to rely on highly optimised and mature MILP solvers in the first place.

Therefore, to obtain robustness for our desired large-scale queries, we must ensure maximum model efficiency, and optimal use of computational resources. While MILP solvers are capable of reliably identifying complex tree structures that elude competing join ordering methods, their use is wasted on queries where ideal solutions are given by straightforward tree shapes. For instance, queries where optimal solutions are given by linear trees can be efficiently solved by conventional approaches such as IKKBZ~\cite{Ibaraki.1984, Krishnamurthy.1986}, while MILP can neither provide any advantage in solution quality, nor match the runtime performance of such efficient baselines. MILP is hence ideally used on precisely those parts of the join tree optimisation where its capability to identify complex, non-linear tree structures is likely to provide the greatest benefit. Conversely, we should rely on more efficient alternatives for the remaining optimisation steps, to maintain MILP model efficiency and thus achieve scalability robustness for large problems. Thus, a \emph{hybrid method} encompassing both, MILP and complementary join ordering methods, is required.

Having outlined the general motivation behind our hybrid method, a series of questions remains to be addressed. In the following, we therefore discuss (a) our selection of a complementary join ordering algorithm, (b) our method to identify suitable cost approximation thresholds, (c) our tree template selection, as well as (d) any necessary MILP model adjustments, before (e) finally discussing our complete hybrid algorithm. 

\subsection{Complementary Algorithm Selection} To select a suitable complementary join ordering approach for our hybrid approach, we firstly consider its required properties. In particular, the algorithm should be characterised by a high algorithmic efficiency, to balance the resource-intensive MILP optimisation step. Conversely, optimising over linear join trees is sufficient for the complementary algorithm, as the tree portions that require more sophisticated bushy shapes are covered by our MILP method. Based on these criteria, we may, for instance, select the conventional IKKBZ algorithm~\cite{Ibaraki.1984, Krishnamurthy.1986}, which identifies ideal left-deep trees in polynomial time. However, when considering further, more recent join ordering approaches, we can identify still more suitable candidates. In particular, Neumann and Radke have proposed a search space linearisation technique~\cite{Neumann.2018}, incorporated into an adaptive framework, to further processes linear trees produced by IKKBZ into more efficient bushy trees using dynamic programming. Their approach is among the most scalable join ordering methods proposed in recent literature, and substantially outclasses IKKBZ solution quality~\cite{Neumann.2018}. We therefore select their adaptive method as the complementary join ordering method to be used alongside MILP in our hybrid framework. For further details on the adaptive method including the search space linearisation technique, we refer the reader to Ref.~\cite{Neumann.2018}.

\subsection{Cost Approximation Thresholds Selection} A further question that remains to be addressed is the selection of cost approximation thresholds. In Example~\ref{example:constr_h}, we have seen how suboptimal threshold selection can severely hamper the optimisation quality of MILP. Yet, an excessively abundant choice of thresholds blows up model size, and curbs optimisation efficiency. Ideally, we can identify a moderate set of precise thresholds marking cost levels of interest, sufficient to closely approximate actual $\mathit{c_{out}}$ costs. To this end, rather than arbitrarily selecting thresholds, we may rely on the solution costs obtained by suitable join ordering alternatives as a baseline for an informed selection. We thus use the adaptive join ordering method~\cite{Neumann.2018}, which is already part of our hybrid method, to swiftly obtain a preliminary solution as a reference for threshold selection. 

As join ordering solutions vary in costs by orders of magnitude, we select thresholds as powers of 2, to capture larger jumps in solution costs. Yet, to retain a lightweight model size, we consider the first power of 2 to exceed overall adaptive costs as an upper bound: Exceeding this threshold implies the current MILP solution fails to improve over the adaptive join tree. We moreover only consider the last few threshold candidates preceding this upper bound, to further reduce model size. In our empirical evaluation, we found a total number of \emph{five thresholds} selected in this manner sufficient to achieve remarkable optimisation quality. 

\subsection{Tree Template Selection}

In addition to approximation threshold selection, the tree template is clearly among the greatest determinants of optimisation quality: Improvements over the existing left-deep MILP approach, or other join ordering methods, are only feasible if the chosen template captures tree shapes approximating optimal join trees. Yet, similarly to approximation thresholds, choosing overly extensive templates results in large MILP models that cannot be optimised efficiently. We therefore require a lightweight template that is likely to encompass optimal solutions. 

To identify suitable template structures, we have quantitatively analysed optimal join trees for queries of small and moderate sizes, where exhaustive search methods like DPSize~\cite{Selinger.1979} can feasibly obtain optimal solutions. In particular, we  assess the relative frequencies of bushy joins within batches of 100 tree queries randomly generated by Neumann and Radke~\cite{Neumann.2018}. Doing so, we obtain insights into effective template structures: While bushy joins are frequently used within the upper depth levels of the join tree, their use becomes increasingly sparse in lower levels of optimal join trees, where optimal solutions tend to constitute linear shapes.

For our method, we therefore select tree templates that encompass balanced tree structures at the top of the join tree, allowing for arbitrary bushy shapes within the first few depth levels. For lower levels, our template may encompass only left-deep structures, or only sparse bushy tree elements. However, given the nature of our method as a hybrid algorithm, we are merely interested in those parts of the template where MILP use can provide an advantage over competitors by identifying complex bushy tree shapes. We therefore apply MILP optimisation exclusively for a limited number of depth levels following the root of the tree, while switching to the more efficient but less exploratory adaptive method for lower levels. Our empirical assessment below will demonstrate the remarkable solution quality achieved by our hybrid method, and thus confirm the aptness of our template selection process.

\subsection{MILP Model Adjustments}
\label{sec:milp_model_adjustments}
As our hybrid method foresees the use of MILP only for limited parts of the join tree, some adjustments are needed for our MILP encoding, which so far requires tree templates to encompass complete join trees, rather than allowing partial join ordering solutions. To render our model compatible with our hybrid algorithm, we seek to specify a set of \emph{anchor joins} $\mathit{AJ}$, corresponding to leaf joins with no further predecessor joins given in the template. For such joins, we must allow the MILP optimiser to terminate without producing complete join trees. 

For each anchor join $\mathit{j} \in \mathit{AJ}$, we introduce a new integer variable $\mathit{nap_j}$  (\emph{Number of Anchor Predecessors}) to our MILP model, where we may freely specify an upper value bound $\mathit{p_{max}}$ marking the maximum number of predecessor joins for $\mathit{j}$, to be later optimised by the complementary adaptive join ordering method. While an increasing number of anchor joins implies more flexibility to escape the rigid bounds of the tree template, the increased flexibility corresponds to an enhanced optimisation complexity. Similarly to the choice of cost thresholds, anchor joins should hence be carefully selected, to maintain a high optimisation quality.

To fully integrate the anchor joins into our MILP model, we must adjust a series of MILP constraints. Firstly, we incorporate them into constraint (A) as $\sum_{j=1}^J \mathit{ja_j} + \sum_{\mathit{j} \in \mathit{JL}}{\mathit{nap_j}} = R-1$, to maintain a total number of $R-1$ actively used joins. Next, we consider constraint (B), which activates joins if any of their direct predecessors have been activated. For each anchor join $j$, we introduce the new constraint variant (B') $\mathit{nap_j} \leq \mathit{p_{max}}\cdot \mathit{ja_{j}}$. Thus, if the optimiser assumes any joins preceding $j$ to be active ($\mathit{nap_j} > 0$), join $j$ must likewise be active to satisfy the constraint in all cases, as $\mathit{nap_j} \leq \mathit{p_{max}}$. Likewise, for constraint (C), which enforces correct numbers of operands for each join, we add the anchor join variant (C') $\sum_r^R \mathit{roj_{rj}} = 2 \cdot \mathit{ja_j} + \mathit{nap_j}$.

\subsection{Hybrid Algorithm}

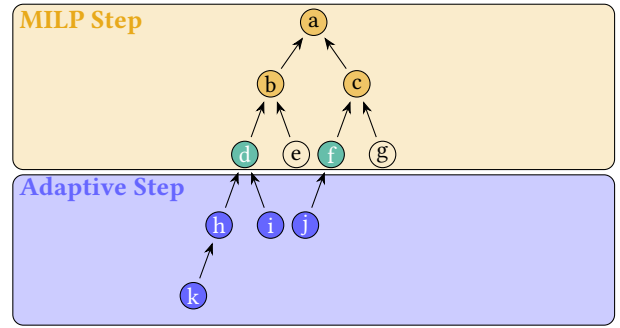
\begin{figure}[htbp]
  \centering
\definecolor{milpcol}{HTML}{E69F00}
\definecolor{leftovercol}{HTML}{009371}
\definecolor{qubitcol3}{HTML}{3468eb}
\definecolor{qubitcol4}{HTML}{6e34eb}

\tikzstyle{adaptive} = [circle, text centered, draw=black, minimum size=10pt, fill=blue!60, inner sep=0pt, outer sep=2pt]
\tikzstyle{milp} = [circle, text centered, draw=black, minimum size=10pt, fill=milpcol!60, inner sep=0pt, outer sep=2pt]
\tikzstyle{unused} = [circle, text centered, draw=black, minimum size=10pt, fill=milpcol!20, inner sep=0pt, outer sep=2pt]
\tikzstyle{leftover} = [circle, text centered, draw=black, minimum size=10pt, fill=leftovercol!60, inner sep=0pt, outer sep=2pt]
\tikzstyle{relationnode} = [circle, text centered, draw=black, minimum size=10pt, fill=teal!60, inner sep=0pt, outer sep=2pt]
\begin{tikzpicture}[node distance=1cm]

\node (nodeMILPBackground)[fill=milpcol!20, rounded
corners, minimum height=2.2cm, minimum width = 8cm, draw=black, yshift=-0.86cm]{};
\node (nodeAdaptiveBackground)[fill=blue!20, rounded
corners, minimum height=2cm, minimum width = 8cm, draw=black, yshift=-3.03cm]{};

\node (milpheader) [minimum width=20, minimum height=0.1cm, rectangle, rounded corners, xshift=-88, yshift=0] {\textcolor{milpcol!90}{\textbf{\large MILP Step}}};

\node (adaptiveheader) [minimum width=20, minimum height=0.1cm, rectangle, rounded corners, xshift=-80, yshift=-63] {\textcolor{blue!60}{\textbf{\large Adaptive Step}}};

    	   \node (node0) [milp] {a};
        \node (node1) [milp]  at ([graphnode, shift=({235:1 cm})]node0) {b};
        \node (node2) [leftover]  at ([graphnode, shift=({250:1 cm})]node1) {\textcolor{white}{d}};
        \node (node3) [unused]  at ([graphnode, shift=({290:1 cm})]node1) {e};
        \node (node5) [milp]  at ([graphnode, shift=({305:1 cm})]node0) {c};
        \node (node4) [leftover]  at ([graphnode, shift=({250:1 cm})]node5) {\textcolor{white}{f}};
        \node (node6) [unused]  at ([graphnode, shift=({290:1 cm})]node5) {g};
        \node (node7) [adaptive]  at ([shift=({250:1 cm})]node2) {\textcolor{white}{h}};
        \node (node8) [adaptive]  at ([shift=({290:1 cm})]node2) {\textcolor{white}{i}};
         \node (node9) [adaptive]  at ([shift=({250:1 cm})]node4) {\textcolor{white}{j}};
         \node (node10) [adaptive]  at ([shift=({250:1 cm})]node7) {\textcolor{white}{k}};

        \draw [reledge] (node1) -- (node0);
        \draw [reledge] (node2) -- (node1);
        \draw [reledge] (node5) -- (node0);
        \draw [reledge] (node3) -- (node1);
        \draw [reledge] (node4) -- (node5);
        \draw [reledge] (node6) -- (node5);

        \draw [reledge] (node7) -- (node2);
        \draw [reledge] (node8) -- (node2);
        \draw [reledge] (node9) -- (node4);
        \draw [reledge] (node10) -- (node7);

        \end{tikzpicture}%
  \captionof{figure}{Illustration of the optimisation steps performed by our hybrid MILP method. Relying on highly optimised MILP solvers, we determine efficient bushy tree structures in the upper tree levels using complete tree templates, with joins a, b, c, d and f selected by the optimiser, while e and g remain inactive. Continuing from the anchor joins d and f (colored in green), we rely on the efficient adaptive algorithm by Neumann and Radke~\cite{Neumann.2018} for lower tree optimisation, yielding joins h-k.}\label{fig:hybrid_overview}
\end{figure}

\SetAlgoNoLine%
\begin{algorithm}[h]
\SetKwFunction{FMain}{HybridMILP}
    \SetKwProg{Fn}{Function}{:}{}
    \Fn{\FMain{$Q=(V,E), \mathit{maxDepth}$}}{
    // Derive adaptive solution (sol) \\
    $\mathit{sol_{adpt}} \gets \mathit{adaptive(Q)}$ \\
    // Derive threshold values for cost approximation \\
    $\mathit{thresholds} \gets \mathit{deriveApproximationThresholds(\mathit{sol_{adpt}})}$ \\
    // Build MILP model for thresholds and max. tree depth\\
    $\mathit{model} \gets \mathit{buildMILPModel(Q, thresholds, maxDepth)}$ \\
    // Perform MILP optimisation \\
    $\mathit{sol} \gets \mathit{optimise(model)}$ \\
    // Derive part. problems from raw MILP solution \\
    $\mathit{partProblems} \gets \mathit{derivePartProblems(\mathit{sol})}$ \\
    // Perform adaptive optimisation on each part. problem \\
    \For{$\mathit{Q_{part}} \in \mathit{partProblems}$}{
    $\mathit{sol_{part}} \gets \mathit{adaptive(Q_{part})}$ \\
    $\mathit{sol.appendPartSolution(sol_{part})}$ \\
    }
    \Return $sol$  \\
}
\caption{Hybrid MILP Method \label{alg:hybrid_MILP}}
\end{algorithm} 

Having outlined all elements of our approach, we now discuss our complete hybrid method as featured in Algorithm~\ref{alg:hybrid_MILP}, and illustrated in Fig.~\ref{fig:hybrid_overview}. Firstly, we rely on the adaptive search space linearisation method by Neumann and Radke~\cite{Neumann.2018} to swiftly obtain a reference solution, and derive our set of cost approximation thresholds based on the adaptive solution costs as an upper cost bound as described above. Next, we build our MILP model for the given query graph $Q$, the approximation thresholds as well as the specified maximum join tree depth to process with MILP. Relying on highly optimised MILP solvers, we obtain a partial solution containing the upper part of the join tree up to the specified maximum depth. Finally, we derive the set of operands associated with each anchor join left incomplete by MILP, which correspond to partial join ordering problems that are efficiently solved using the adaptive algorithm. Each respective partial join tree solution is then appended to the upper join tree portion obtained by MILP, to obtain the complete join tree solution. 

\section{Experimental Analysis}
\label{sec:experimental_analysis}
In this section, we experimentally verify the aptness of our novel hybrid MILP method for large-scale join order optimisation. We discuss our experimental setup in Sec.~\ref{sec:experimental_setup}, present results for conventional query optimisation benchmarks in Sec.~\ref{sec:conventional_benchmarks}, and finally assess scalability to extremely large query sizes in Sec.~\ref{sec:scalability_analysis}.

\subsection{Experimental Setup}
\label{sec:experimental_setup}

In our analysis, we seek to identify the most suitable method for large-scale join order optimisation. To this end, we seek to include a wide range of methods with varying properties, prompting the following selection of baseline algorithms:

\begin{itemize}[nosep,left=0pt]
    \item DPSize: \emph{Dynamic programming~(DP)} method, building optimal bushy join trees without cross products~\cite{Selinger.1979}.
    \item DPHyp: \emph{Dynamic programming~(DP)} method, featuring improved algorithmic efficiency over DPSize~\cite{Moerkotte.2008}.
    \item IKKBZ: \emph{Polynomial-time}  algorithm, yielding optimal \emph{left-deep trees} for \emph{acyclic query graphs}~\cite{Ibaraki.1984, Krishnamurthy.1986}.
    \item Adaptive: \emph{Adaptively selects algorithms} based on query graph size and properties. For problems considered in our paper, the method applies a \emph{search space linearisation technique} refining IKKBZ solutions into bushy trees via dynamic programming~(linearizedDP)~\cite{Neumann.2018}.
    \item Greedy Operator Ordering~(GOO): \emph{Greedy bottom-up} heuristic yielding \emph{bushy} trees~\cite{Fegaras.1998}.
    \item GOO-DP: Algorithm \emph{refining GOO solutions} via dynamic programming~\cite{Neumann.2018}.
    \item Minsel: \emph{Greedy algorithm} yielding left-deep trees~\cite{Swami.1989}.
    \item Genetic: \emph{Genetic algorithm} yielding bushy trees. 
    \item QuickPick: \emph{Randomised algorithm} yielding \emph{bushy trees}~\cite{Waas.2000}.
    \item Simplification: Heuristic that \emph{greedily prunes} the query graph, obtaining \emph{bushy trees}~\cite{Neumann.2009}.
   
\end{itemize}

Our selection of algorithms covers a wide range of paradigms, including (1) dynamic programming methods like DPSize and DPHyp, as conventionally applied by query optimisers to ensure optimal solutions, (2) polynomial-time methods like IKKBZ, guaranteeing optimal solutions in a reduced linear search space, (3) greedy heuristics such as Minsel or GOO, trading solution quality for algorithmic efficiency, (4) probabilistic methods like QuickPick, as well as (5) genetic algorithms. We further include the adaptive method proposed by Neumann and Radke~\cite{Neumann.2018}, which applies a search space linearisation based on IKKBZ result for the problem sizes and properties considered in our analysis. Their approach is among the most robust methods for large-scale query optimisation proposed in the recent literature, and thus constitutes one of the most suitable baselines to compare against. In our analysis, we will show how our hybrid MILP method, which incorporates the adaptive algorithm as discussed in Sec.~\ref{sec:hybrid_algorithm}, improves over the standalone adaptive method, as well as the remaining wide range of considered join ordering competitors.

For all baseline algorithms, we use implementations by Neumann and Radke~\cite{Neumann.2018}. For our MILP method, we use our own implementation in Python (version 3.10.14), using the conventional Gurobi MILP solver with the gurobipy package (version 12.0.2)~\cite{gurobi}. To create our MILP model, our hybrid algorithm as featured in Algorithm~\ref{alg:hybrid_MILP} builds a tree template containing every join up until the specified $\mathit{maxDepth}$ parameter. We run four configurations with depths $(4, 5, 6, 7)$, which we find sufficient to determine optimal bushy tree structures. We consider the best result obtained by all configurations, which correspond to individual MILP models that can be optimised in parallel. Further, we specify two anchor joins as described in Sec.~\ref{sec:milp_model_adjustments}: For both, the left and right sub tree joined by the root join, anchor joins correspond to the left-deep join of either tree half, as illustrated in Fig.~\ref{fig:hybrid_overview}. Our implementation is provided in our \suppweb~\cite{Mauerer:2021}. 
We run all experiments on a system with two AMD EPYC 7662 CPUs and 1 TB of RAM. All algorithms are set to time out after 60 seconds. 

Finally, rather than raw queries, the input to the join ordering problem is given by a query graph as discussed in Sec.~\ref{sec:join_ordering_model}. Accordingly, for all queries considered in our analysis, we pass query graphs as extracted by Neumann and Radke~\cite{Neumann.2018} for both, the conventional benchmark and synthetic tree queries that we analyse in the following, to each algorithm.

\subsection{Conventional Benchmarks} 
\label{sec:conventional_benchmarks}

We begin our analysis by considering conventional benchmarks, including TPC-H~\cite{tpc-h}, TPC-DS~\cite{tpc-ds}, LDBC BI~\cite{Angles.2014}, as well as the join ordering benchmark~(JOB)~\cite{Leis.2018}. For the query sizes featured in these benchmarks, exhaustive search methods such as DPSize~\cite{Selinger.1979} obtain optimal solutions within their considered solution space well within our 60s time limit. For these benchmarks, we therefore restrict our analysis to comparing MILP solution quality against the DPSize method, which yields an upper bound on the solution quality for all considered competing join ordering methods. Further below, we will contrast the full set of algorithms for substantially larger tree queries, where performance differences between individual methods become significantly more pronounced. 

For all benchmark queries, Neumann and Radke~\cite{Neumann.2018} extracted their corresponding query graphs, which we consider in our analysis. We restrict the problem set to those queries conforming to the conventional join ordering model outlined in Sec.~\ref{sec:join_ordering_model}. Since all queries featured by the benchmarks are of only small sizes, our hybrid framework is not required to maintain performance for large problems, and we can instead rely on our pure, standalone MILP method as presented in Sec.~\ref{sec:milp_encoding}. We apply cost approximation thresholds as powers of 2, and use a tree template featuring joins to allow for both, balanced and left-deep join trees, as well as any trees interpolating between these shapes. Code for our template generation can be found in our~\suppweb.

\begin{figure}[htbp]
  \includegraphics{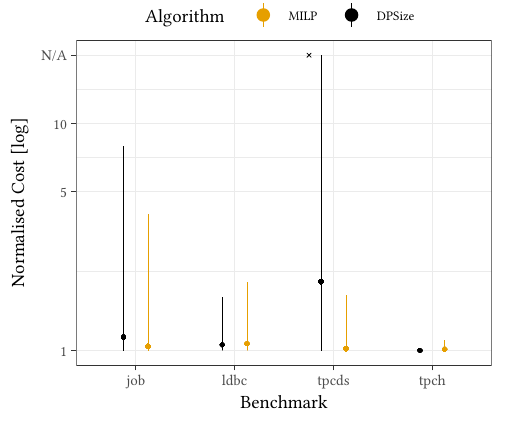}\vspace*{-1em}
  \caption{Normalised solution costs (relative to overall best solutions) for conventional benchmark queries. Range lines visualise minimal, average and maximal normalised costs over individual queries. Crosses with value N/A represent individual solutions with prohibitively large costs \(\geq\) 20.}\label{fig:benchmark_results}
\end{figure}

Fig.~\ref{fig:benchmark_results} contrasts the solution quality achieved by our MILP method against DPSize solutions. We depict normalised solution costs, relative to the best solution obtained by any algorithm, as cost ranges indicating minimal, average and maximal solution costs over all queries featured by each respective benchmark (notice that we do not use boxplots that would degenerate to to a single line, as the results are sharply centered around the mean value). We consider normalised solution costs of 20 as an upper bound for visualisation.

While DPSize obtains optimal solutions within its considered solution space, its exhaustive search only explores solutions that do not rely on cross product operations. In contrast, our MILP model considers the complete search space including cross products. Thereby, the MILP solver often beats DPSize solution quality: For the JOB, average normalised costs are given by 1.04 for MILP and 1.14 for DPSize, with maximum MILP costs of 4 compared to maximum DPSize costs of 7.97. The performance gap is still more pronounced for TPC-DS, with average DPSize costs of 2 compared to average MILP costs of 1.01, and maximum DPSize costs exceeding our normalised cost bound of 20.

Our MILP method consistently obtains optimal or near-optimal solutions for almost all of the 282 queries collectively featured by all benchmarks, exceeding normalised costs of 2 only for two particular queries\footnote{While normalised MILP worst-case costs are given by 4, they occur for a query with minimal absolute costs of 1. Thus, even the slightly suboptimal MILP solution with absolute costs of 4 results in a big normalised cost value.}. By identifying beneficial cross product operations, our approach significantly improves over DPSize solution quality, which exceeds normalised costs of 2 for 23 benchmark queries, and which moreover constitutes an upper bound on solution quality for the remaining competitors listed above.  

\subsection{Scalability Analysis}
\label{sec:scalability_analysis}

\begin{figure*}[htbp]

  \includegraphics{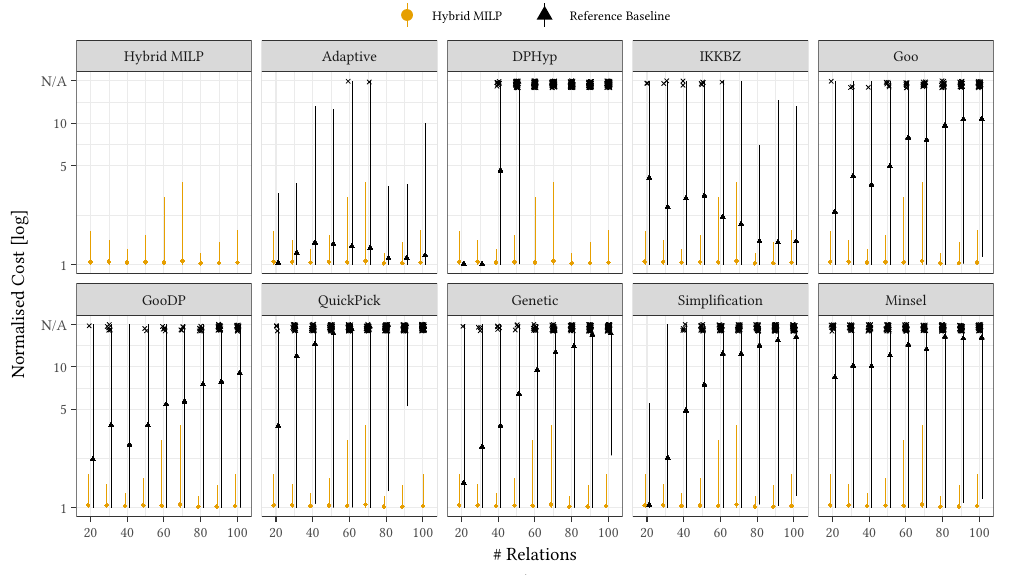}\vspace*{-1em}
  \caption{Normalised solution costs (relative to overall best solutions) for tree queries joining increasing numbers of relations. Range lines visualise minimal, average and maximal normalised costs for a set of 100 problem instances. Crosses with value N/A (slightly jittered horizontally and vertically to resolve outlier count) represent individual solutions with prohibitively large costs \(\geq\) 20, or scenarios where an algorithm failed to obtain any result within our 60s time limit.}\label{fig:tree_results}
\end{figure*}

While our analysis for conventional benchmarks indicates the robust performance of our MILP method, the sizes of queries considered so far remain insufficient to assess its scalability aptness\footnote{The biggest queries among all considered conventional benchmarks join 18 relations.}. As such, we now consider tree queries generated by Neumann and Radke~\cite{Neumann.2018} to benchmark methods for large-scale join order optimisation. Fig.~\ref{fig:tree_results} depicts normalised solution costs for 900 tree queries joining up to 100 relations. Each depicted query size features 100 individual queries. 

In the following, we discuss the performance of each considered join ordering algorithm in detail.

\paragraph{DPHyp}
We begin with DPHyp, which applies dynamic programming to guarantee optimal bushy join trees without cross products. For small and moderate queries, the method thereby beats all other methods considered in our analysis, which regularly yield at least slightly suboptimal results. However, the biggest drawback of the method becomes apparent once queries further grow in size: For 40 relations, DPHyp starts experiencing occasional timeouts, and beginning at 50 relations, the method fails to yield any result within 60s for all but two of the 100 randomly generated tree queries~\footnote{Note that our visualisation in Fig.~\ref{fig:tree_results} maps scenarios with timeouts to our upper "N/A" cost bound of 20. This prompts average DPHyp costs to exceed 1 once the method fails to obtain optimal solutions within our 60s time limit.}.
To achieve robustness not only for small queries, but even within extremely large solution spaces, we hence require suitable alternatives. 

\paragraph{IKKBZ}
Similarly to DPHyp, the conventional IKKBZ algorithm provides guarantees on solution quality; yet, its polynomial-time characteristics render it substantially more efficient, avoiding the timeout issues of DPHyp. However, while the method promises to yield optimal solutions, it is restricted to a search space strictly containing left-deep join trees. While such trees often correspond to, or approximate optimal tree structures, failure to capture bushy tree structures commonly results in a blow-up of solution costs, as shown by our empirical results: Worst-case costs exceed our upper normalised cost bound of 20 for most problem sizes, while even the average normalised IKKBZ costs reach as high as 4.1 for 20 relations. Identifying scalable join ordering methods to obtain efficient plans encompassing bushy join trees thus remains desirable. 

\paragraph{Minsel}
Like IKKBZ, the Minsel algorithm only considers a left-deep solution space, but applies a greedy heuristic to build join trees without any guarantees on solution quality. Accordingly, Minsel is among the algorithms yielding the lowest solution quality, with costs often far exceeding even our generous normalised cost bound of 20, by orders of magnitude. While this unsteady performance is characteristic for greedy optimisation approaches, it is further reinforced by the method's inability to account for bushy tree structures. We thus next consider further refined greedy approaches.

\paragraph{GOO} 
In contrast to Minsel, the GOO method applies a greedy search strategy to build join trees that can feature bushy tree structures. The method thereby achieves a significantly more robust performance compared to Minsel. However, it struggles to approximate the solution quality of other methods considered in our analysis, and, like Minsel, exceeds our normalised cost bound of 20 for problems of all considered query sizes. Average GOO costs start at 2.36 for 20 relations, but grow as high as 10.7 for our largest problem class featuring 100 relations, corresponding to very inconsistent performance.

\paragraph{GOO-DP}
To boost GOO solution quality, the GOO-DP method applies dynamic programming to further refine GOO join trees. Hence, GOO solutions provide an upper bound on GOO-DP solution costs. While the method does improve GOO solution quality, cost improvements are only moderate, and even the refined GOO-DP solutions fail to match the quality of competitors considered in our analysis: Like standard GOO, the method often exceeds our normalised cost bound for all problem sizes, with average GOO-DP costs reaching up to 8.99 for our largest class of 100 relations. 

\paragraph{Simplification} We next consider the simplification method as a final algorithm relying on greedy optimisation steps. For our smallest classes of join ordering problems, the method tends to identify significantly better solutions compared to the other greedy algorithms discussed so far: For 20 relations, average costs are close to the optimal costs of 1, as compared to 2.35 and 2.2 for GOO and GOO-DP respectively, while growing to a worst case value of 5.5 for one particular problem. Yet, the simplification method scales less gracefully, leading to a quick deterioration of solution quality once query sizes grow. Starting at 40 relations, the method fails to beat GOO and GOO-DP in the average case, with worst case costs well beyond our normalised cost bound of 20. In conclusion, none of the greedy algorithms considered in our analysis manage to achieve scalability robustness for large query sizes.

\paragraph{QuickPick} Rather than applying dynamic programming or greedy optimisation steps, QuickPick, as implemented by Neumann and Radke~\cite{Neumann.2018}, randomly generates \num{1000} join trees, and picks the best solution found in the process. For a small number of problem scenarios, this search flexibility enables the method to obtain optimal solutions that elude most other join ordering methods considered in our analysis. However, the drawbacks of randomised methods like QuickPick quickly become apparent when considering their overall performance, which is subpar compared to most other algorithms: Even for 30 relations, average normalised costs exceed 10, whereas most of the QuickPick solutions reach our upper cost bound of 20 starting at 60 relations. QuickPick is hence among the most unreliable methods considered in our analysis, alongside the Minsel algorithm. 

\paragraph{Genetic} We next consider the genetic algorithm as a further randomised metaheuristic, which is applied, for instance, by the PostgreSQL query optimiser for queries joining at least 12 tables. Similarly to QuickPick, its rather flexible solution space exploration enables the genetic algorithm to identify efficient bushy trees in a few cases where more rigid competitors fail to approximate the optimal tree structure. Yet, while genetic performance is significantly more stable compared to QuickPick, with lower average solution costs in all scenarios, solution quality still degrades substantially once queries grow in size. Starting at 50 relations, average costs exceed 5, and approximate our normalised cost bound of 20 once queries join 100 relations. Much like the greedy algorithms, randomised approaches like QuickPick or the genetic algorithms thus fail to yield a sufficient level of robustness for large-scale query optimisation. 

\paragraph{Adaptive} For the problem sizes and characteristics considered in our analysis, the adaptive algorithm applies a search space linearisation method to refine linear IKKBZ solutions into bushy trees, using dynamic programming. Thereby, the method inherits the algorithmic efficiency of IKKBZ, and overcomes its linear search space limitation. Doing so, the method significantly improves over all competitors considered so far: Average normalised adaptive costs tend to be near-optimal for all problem sizes. While these empirical results confirm the overall effectiveness of the search space linearisation, the guarantees on solution quality provided by IKKBZ for linear solutions do not directly translate to the linearised bushy trees. In particular, as the solution range explored by the method is confined by the IKKBZ baseline, the method fails to identify optimal trees that diverge from the linearised solution space. This results in suboptimal worst-case behavior: For 40 and 50 relations, worst-case adaptive costs exceed normalised optimal costs by a factor larger than 10. For 60 and 70 relations, some adaptive solutions even exceed our normalised cost bound of 20. Finally, these worst-case characteristics hold up to our largest query sizes joining 100 relations, with worst-case adaptive costs of 10. While the adaptive method efficiently yields computationally cheap plans in cases where optimal trees conform to the linearised solution space, its worst-case performance results in suboptimal plans exceeding optimal solution costs by orders of magnitude.

\paragraph{Hybrid MILP} To address the limitations of the join ordering methods discussed so far, we finally consider our novel hybrid MILP method: 
By applying a flexible search space exploration using MILP, our method identifies efficient bushy tree structures at the tree top, while maintaining algorithmic efficiency by swiftly optimising lower tree parts using the adaptive method. Thereby, our method overcomes the drawbacks and limitations of competing approaches: The hybrid MILP method obtains near-optimal solutions in almost all scenarios, while avoiding the suboptimal worst-case performance of the adaptive method and other competitors. Among the 900 tree queries considered in our analysis, normalised costs of 2 are exceeded in merely two cases (with maximum costs of 3.86) by our hybrid MILP method, as compared to 47 cases for the adaptive method as its closest competitor. This demonstrates the remarkable robustness of our hybrid MILP method for large-scale query optimisation.

So far, we have only considered the quality of solutions in our analysis. We complete our analysis by considering the runtime behavior of our method compared to competitors. While our hybrid MILP method obtains solutions well before our 60s timeout, the resource-intensive nature of MILP optimisation renders approaches relying on MILP less time-efficient by default compared to most of the competitors considered in our analysis, with the exception of dynamic programming methods such as DPHyp. While optimisation times for other competitors range in the milliseconds, MILP requires up to 23 seconds for the most complex queries until solution quality converges: Table~\ref{tab:milp_times} details the average optimisation times required until solution costs are within a 20\% threshold of the final MILP result, for increasing numbers of relations. Note that the convergence times tend to, but do not need to strictly increase as queries grow in size, as particular smaller queries may be more complex to optimise compared to larger ones.

While the runtime behavior of MILP approaches cannot match most of the considered competitors, our empirical analysis has shown that relying on mature MILP solvers and investing increased computational resources allows us to substantially improve over the solution quality of most competitors. By avoiding the worst-case costs of competing methods, which exceed our hybrid MILP plan costs by orders of magnitude, our method yields efficient plans in all of the scenarios considered in our paper, and thus provides a robust, novel alternative for large-scale query optimisation.

\begin{table}[htbp]
	\centering
	\caption{Average MILP convergence times for increasing numbers of relations. We measure MILP result quality in intervals of 10s. Accordingly, our reported values capture upper bounds on convergence times, that may further decrease with more fine-grained interval measurements.}
	\begin{tabular}{lrrrrrrrr}  
		\toprule
		\# Relations & 30 & 40 & 50 & 60 & 70 & 80 & 90 & 100 \\
		\midrule
	 Time [s] & 10 & 10.8 & 14.2 & 16.1 & 22.6 & 15.3 & 20.3 & 18.6 \\
     \bottomrule
	\label{tab:milp_times}
	\end{tabular}\vspace*{-2em}
\end{table}

\section{Related Work}
\label{sec:related_work}
Join ordering algorithms can be roughly divided into two distinct categories. The first class of approaches comprises exhaustive search methods that seek out optimal solutions relying on techniques such as dynamic programming. Representatives of this category include the conventional DPSize method by Selinger \etal~\cite{Selinger.1979}, as well as the more efficient DPHyp algorithm by Moerkotte and Neumann~\cite{Moerkotte.2008} assessed in our empirical analysis, in addition to a series of further methods proposed in the literature~\cite{graefe1993volcano, Meister.2020, Moerkotte.2006, Vance.1996, Moerkotte.2008}. While such join ordering approaches provide formal guarantees on solution quality, the challenging growth of the join ordering solution space restricts their use to queries of small and moderate sizes. We have empirically shown their limits in our analysis, demonstrating the need for suitable alternatives for large-scale join order optimisation.

To efficiently process larger queries, query optimisers switch from exhaustive search to heuristic methods, which constitute the second broad category of join ordering approaches. They include a large variety of methods, such as greedy algorithms~\cite{Fegaras.1998, Swami.1989, Neumann.2009}, genetic algorithms~\cite{Steinbrunn.1997, Horng.1994} and more general probabilistic approaches~\cite{Waas.2000}, among others~\cite{Bruno.2010, Ioannidis.1990, Swami.1988, Trummer.2016b}. While heuristic methods can maintain algorithmic efficiency even for large-sized queries, the quality of heuristic solutions tends to degrade as queries grow in size, as outlined by our experimental results. This prompts highly sub-optimal plans, whose costs exceed optimal solution costs by orders of magnitude, resulting in large computational overheads during query execution. Thus, more robust methods are required to address large-scale queries. 

A prominent technique to reduce the exploration complexity is given by solution space pruning, as applied by the conventional IKKBZ method~\cite{Ibaraki.1984, Krishnamurthy.1986}. By restricting the search to the subset of linear join ordering solutions, the polynomial-time IKKBZ method efficiently yields optimal linear solutions. Yet, the restriction to linear join trees frequently yields highly sub-optimal solutions compared to general bushy join trees. To improve the solution quality in such cases, Neumann and Radke~\cite{Neumann.2018} have proposed a search space linearisation technique that transforms linear IKKBZ solutions into bushy trees via dynamic programming. However, linearised plans remain costly if the linearisation technique fails to approximate the ideal tree structure, prompting highly sub-optimal worst-case behaviour as observed in our empirical analysis. 

Similar limitations apply to the existing MILP method for join order optimisation proposed by Trummer and Koch~\cite{Trummer.2017}: As their model only explores a left-deep solution space, plan costs can substantially exceed optimal solutions. In contrast, we have proposed a novel MILP encoding capable of exploring bushy tree structures as defined by an arbitrary tree template. By selecting suitable templates, and by combining our MILP method with complementary join ordering methods, our method overcomes the limitations of the existing left-deep MILP approach, and avoids the worst-case behaviour of other join ordering algorithms discussed above.
Our approach connects to a series of special-purpose optimisation methods recently proposed for query optimisation, which not only include the existing MILP method~\cite{Trummer.2017}, but also further constraint optimisation approaches of varying kinds. They include, for instance, methods relying on quantum computing devices~\cite{Schoenberger.2022, Schoenberger:2023, schoenberger:23:qdsm, nayak.2023, franz:2024}, which requires optimisation models where constraints are implicitly encoded into a unified cost formula~\cite{gogeissl:24:qdsm,schmidbauer:24}. While contemporary quantum systems are mere prototypes~\cite{Greiwe:2023,Safi:2023}, and hence incapable of large-scale optimisation, Schönberger~\etal have demonstrated the use of quantum-inspired high-performance systems like the Fujitsu Digital Annealer on larger queries~\cite{schoenberger:23:pvldb, Schoenberger:25sigmod}. Yet, similarly to the existing MILP method by Trummer and Koch~\cite{Trummer.2017}, their method is limited to a left-deep solution space, which degrades plan quality in many scenarios.

\section{Discussion and Conclusion}
\label{sec:conclusion}
Join order optimisation remains one of the most relevant problems in query optimisation, and the broader domain of data management. While query optimisers can rely on exhaustive search to obtain ideal join trees for small queries, such methods become infeasible for larger problems, given the extreme growth of join ordering solution spaces. Yet, large queries have become increasingly frequent in real-world scenarios, and require adequate means of processing.

Despite decades of research, finding optimal solutions for large query loads remains challenging, as shown by our empirical analysis: The quality of plans produced by typical heuristics (such as greedy optimisation, probabilistic methods, or genetic algorithms) worsens with growing queries, with heuristic plan costs exceeding optimal solution costs by orders of magnitude. While search space linearisation techniques achieve more robust performance by refining optimal linear solutions into bushy trees, their solution quality degrades for queries where the linearisation fails to capture the truly optimal bushy tree structures, resulting in highly suboptimal worst-case behaviour across all analysed query classes.

To address limitations of these approaches, we developed a novel join ordering method that relies on highly efficient MILP solvers, which have matured over decades. Despite their remarkable performance in a wide range of domains, they remain underutilised in query optimisation. Improving over the existing MILP model proposed for join ordering, which was restricted to left-deep join trees, we have derived a novel MILP encoding that allows the optimisation of arbitrary join tree structures. By embedding our MILP method into a hybrid framework, we apply MILP solvers precisely where they provide the biggest advantage over competing methods, while switching to more efficient, yet less exploratory join ordering methods for the remaining solution portions.

Among the wide range of join ordering methods assessed in our paper, our hybrid MILP approach thereby achieves the most robust performance for large-scale join order optimisation: Our method consistently obtains optimal or near-optimal solutions for NP-hard tree queries joining up to 100 relations, which far exceeds typical query sizes. By relying on optimised MILP solvers, our method avoids highly suboptimal plans resulting from the worst-case behaviour of competitors, and thus constitutes a novel, highly robust alternative for large-scale join order optimisation. Our results outline the potential of special-purpose solvers for query optimisation, and prompt the further exploration of MILP and further constraint optimisation methods for still unconsidered problems in the data management domain.

\begin{acks}
WM acknowledges support by the High-Tech Agenda Bavaria.
\end{acks}

\bibliographystyle{ACM-Reference-Format}
\bibliography{literature}

\end{document}